\documentclass[
 amsmath,amssymb,
 aps,
pra,twocolumn, superscriptaddress]{revtex4-2}

\usepackage{graphicx}
\usepackage{dcolumn}
\usepackage{bm}
\usepackage{xcolor}
\usepackage[colorlinks= true,urlcolor=blue,linkcolor= blue,citecolor=blue,bookmarks=false,pdfstartview=]{hyperref}

\begin{document}



\title{Optically-induced electric Aharonov-Bohm effect}


\author{Gabriel P. Martins}
\affiliation{Departamento de Física, Universidade Federal de Minas Gerais, Belo Horizonte, Minas Gerais 31270-901, Brazil}
\author{Herman Batelaan}
\affiliation{Department of Physics and Astronomy, University of Nebraska—Lincoln, 208 Jorgensen Hall, Lincoln, Nebraska 68588-0299, USA}
\author{Pablo L. Saldanha}\email{saldanha@fisica.ufmg.br}
\affiliation{Departamento de Física, Universidade Federal de Minas Gerais, Belo Horizonte, Minas Gerais 31270-901, Brazil}

\date{\today}

\begin{abstract}
Although it was proposed almost 70 years ago, the electric version of the Aharonov-Bohm (AB) effect, where the interference pattern of quantum charged particles is affected by electric fields that do not interact locally with them, was never experimentally implemented. Here we propose a feasible way to experimentally verify such an important effect, for an electron double-slit interferometer with the slits carved in a nonlinear crystal. The AB phase is generated by a quasi-DC electric field induced in the crystal with the passage of a light pulse (the optical rectification effect) while the electron is superposed in the two slits. We show that this procedure can change the interference pattern of a ensemble of electrons even if they are subjected to negligible electric fields during their propagation, considering feasible parameters.
\end{abstract}


\maketitle


\section{Introduction} \label{intro}

In the Aharonov-Bohm (AB) effect, an interferometer for a quantum charged particle with a non-zero electromagnetic field in the region in between its paths exhibits an interference pattern that depends on the  field, even if the field on the paths themselves is identically zero \cite{Ehrenberg:1949mei,aharonov59,peshkin05}. The physical interpretation of this phenomenon is the subject of ample discussions over the last decades. Originally, it was interpreted as the result of local interactions between the quantum particle and the electromagnetic potentials \cite{aharonov59}, which would give physical reality to the potentials, previously believed to be mere mathematical tools. 
But it can also be interpreted as a phase shift in the superposed fields of the quantum particle and the electromagnetic field \cite{Peshkin:1981hz}, or as a phase shift accumulated by the charged particles that are the source of the electromagnetic field moving in the quantum particle’s potential \cite{Vaidman2012,Pearle:2017swr,Pearle:2017whb}, which would affect the quantum particle behavior in a nonlocal way. In a quantum treatment for the electromagnetic field, the AB effect can be interpreted as the result of an exchange of virtual photons between the quantum particle and the source of the potentials \cite{Santos1999,Marletto2020,Saldanha2021,Saldanha:2021,Saldanha:2024vid}.

The AB effect is usually separated into two categories. If there is only a magnetic field in between the interferometer paths, it is called ``magnetic AB effect". On the other hand, if only an electric field is present, it is called ``electric AB effect". But it is worth mentioning that both electric and magnetic fields may be present in the interferometer, as in the electrodynamic AB effect \cite{Saldanha2023,Saldanha:2025}, for instance. And with a change of reference frame, a magnetic AB effect can turn to an electric AB effect and vice-versa \cite{Rodrigues2025}. The magnetic AB effect has been observed in multiple settings over the past decades \cite{chambers60,tonomura82,webb85,tonomura86,bachtold99,ji03,peng10,becker19,nakamura19,deprez21,ronen21}. 
{The electric AB effect, on the other hand, has proven to be very difficult to achieve. Electric AB phase shifts have been experimentally observed in configurations where the charged particles are subjected to local electric fields along their propagation through the interferometer \cite{Matteucci1985,vanOudenaarden1998}, allowing the experimental results to be interpreted in terms of such local interactions. In particular, the experiment of van Oudenaarden et al. \cite{vanOudenaarden1998}, sometimes referred to as an observation of the electric AB effect, involved electrons traversing regions with electric fields generated by the applied bias voltage. The experimental observation of the electric AB effect for free electrons propagating through regions with negligible electric fields has not yet been demonstrated.}

In this work, we present the theoretical background for an experimentally achievable, optically-induced electric AB effect. In our proposal, a laser pulse passes in between the holes of a double slit electron interferometer. The slits are carved in a nonlinear optical crystal. Such a crystal is more easily polarizable in one direction  than in the opposite one \cite{boyd20}, which leads to a non-zero net average electric field inside the material during the passage of the laser pulse, in the so-called optical rectification effect. An electron which is in a superposition state in the two slits when the laser pulse passes through will experience a negligible net force due to residual fields, and an AB phase difference will be created between the two possible paths, due to the nonzero average electric field inside the crystal, or alternatively, due to the local interaction of the
electron with the scalar potential. We discuss that such an experiment can be performed with current technology.



\section{Creating DC electric fields with light in a nonlinear crystal} \label{nonlinear}

The ability to engineer and manipulate electronic phases has become an important tool in electron interferometry and coherent electron optics~\cite{batelaan07,barwick09}. Optical techniques are especially attractive in this context due to their dynamical controllability and ultrafast timescales~\cite{feist15,vanacore18}. In most realizations, however, optically induced phase shifts originate from direct local interactions between electrons and electromagnetic waves along their trajectories~\cite{feist15,vanacore18,echternkamp16}. Here, to produce an electric AB scheme where the charged quantum particles propagate only where the electric field is zero, we propose to use the optical rectification effect to produce an electric field which is negligible at the possible positions of the particles, but which induces an AB phase difference between the interferometer paths. Being optically induced, this electric field can last for a very short time.

\begin{figure}
\includegraphics[width = 8.5cm]{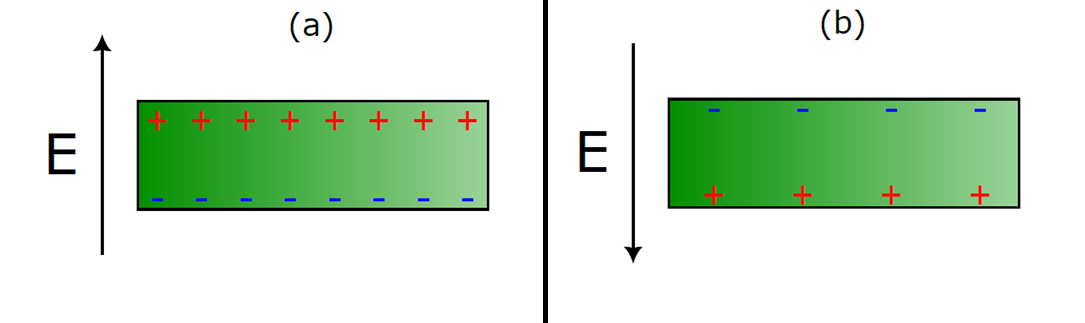}
\caption{The response of a nonlinear crystal to $y$-polarized electric fields of equal intensities, but in opposite directions. The non-linearity of the crystal's susceptibility is such that it favors polarization when the $E$-field is in the positive $y$-direction (a), rather than when it is in the negative $y$-direction (b).\label{schem}}
\end{figure}

When a crystal with non-linear electric susceptibility is subjected to an electric field $ \mathbf{E}_\omega(\mathbf{r})e^{-i\omega t}$,
the resulting electric polarization $\mathbf{P}$ can be decomposed into components with several frequencies. Up to second order, the polarization field is given by
\begin{equation}
    \mathbf{P}(\mathbf{r},t) = \mathbf{P}_\omega(\mathbf{r}) e^{-i\omega t} + \mathbf{P}_{2\omega}(\mathbf{r})e^{-2i\omega t} + \mathbf{P}_{OR}(\mathbf{r}),
\end{equation}
where $\mathbf{P}_\omega(\mathbf{r}) = \varepsilon_0\chi^{(1)}\mathbf{E}_\omega(\mathbf{r})$, is the first order term; $[\mathbf{P}_{2\omega}]_i(\mathbf{r}) = \varepsilon_0\sum_{jk}\chi^{(2)}_{ijk}(2\omega,\omega,\omega)[\mathbf{E}_\omega(\mathbf{r})]_j[\mathbf{E}_\omega(\mathbf{r})]_k$, is the second order term responsible for the appearence of second harmonic photons; and $[\mathbf{P}_{OR}]_i(\mathbf{r}) = \varepsilon_0\sum_{jk}\chi^{(2)}_{ijk}(0,\omega,-\omega)[\mathbf{E}_\omega(\mathbf{r})]_j[\mathbf{E}^*_\omega(\mathbf{r})]_k$ is the optically rectified polarization field \cite{boyd20}. The appearance of a non-zero average polarization $\mathbf{P}_{OR}$ is due to the fact that some non-linear crystals are more easily polarized in one direction than in the other, as depicted in  Fig. \ref{schem}. Throughout the remainder of this paper, we will be studying the optically rectified field generated by external electric fields linearly polarized in the direction of the crystal's slow axis. In this case, we can rewrite $\mathbf{P}_{OR}(\mathbf{r})$ in a simpler notation as 
\begin{equation}
\mathbf{P}_{OR} = \varepsilon_0\chi^{(2)} |{\mathbf{E}_{\omega}}|^2 \hat{\mathbf{y}}, \label{ORfield}
\end{equation}
where we have taken the $\hat{\mathbf{y}}$ direction to be that of the slow axis, with $\chi^{(2)}\equiv\chi^{(2)}_{yyy}(0,\omega,-\omega)$.

The polarization field $\mathbf{P}$ generates bulk ($\rho_p$) and surface ($\sigma_p$) polarization charges in the crystal, given by $\rho_p = -\nabla\cdot\mathbf{P}$, and $\sigma_p = \mathbf{P}\cdot\hat{\mathbf{n}}$, where $\hat{\mathbf{n}}$ is the vector normal to the crystal edges. One can then use these polarization charge densities to calculate the electric field inside the crystal and in its vicinities using Coulomb's law. In order to calculate the average field, one needs only to consider the average optically rectified polarization field $\mathbf{P}_{OR}$, since the other terms average to zero.

\section{Proposal to verify the electric AB effect in a double-slit interferometer} \label{double-slit}

In our proposal for an optically induced electric AB effect, we consider a slab of a nonlinear crystal with a double slit carved into it. {For the numerical estimates presented below, we consider potassium
titanyl phosphate (KTP) as the nonlinear crystal, with
$\chi^{(2)} \approx 33\times10^{-12}\,\mathrm{m/V}$ and refractive
index $n=1.8$.} Electrons pass through the double slit and form an interference pattern in a screen placed in front of it.  As we discuss in the following, a light pulse can induce charge densities in the slab due to the optical rectification effect while each electron is in a superposition state in both holes. These charges can generate a potential difference between the two paths, changing the interference pattern even if the electrons only propagate in regions with negligible electric fields.

\begin{figure}
    \includegraphics[width=0.4\textwidth]{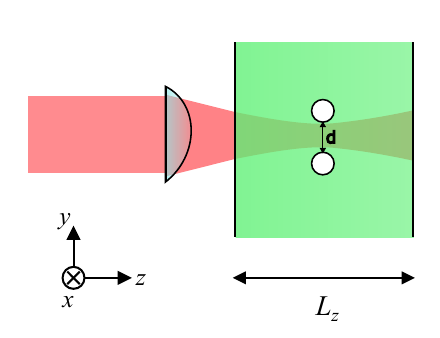}
\caption{A gaussian, linearly-polarized laser beam is collimated by a  cylindrical lens. The collimated beam then enters a non-linear crystal slab with two circular slits carved into-it. The collimated laser passes through in between the circular slits separated by a distance $d$. The width of the crystal is $L_x$ in the $x$-direction, and $L_z$ in the $z$-direction, and the separation between the two circular holes is $d$. \label{setup}}
\end{figure}

As depicted in Fig. \ref{setup}, two circular holes separated by a distance $d$ are carved in the nonlinear crystal slab.  A  pulsed  gaussian laser beam, propagating in the $z$-direction, is collimated   by a cylindrical lens in a way that the beam waist fits in between the holes. The collimated beam is such that it is narrowly collimated in the $y$-direction, so that it fits between the holes, but it is still large in the $x$-direction, much larger than the crystal width in the $x$-direction $L_x$. So, we may consider that the field amplitude does not vary with $x$ inside the crystal. During the laser pulses, the DC polarization inside the slab will be given by Eq. (\ref{ORfield}), where we consider that the amplitude of the external gaussian beam field is given by 
\begin{equation}\label{Elaser1}
    |\mathbf{E}_{\omega}| = |E(z)| \exp\left[\dfrac{-y^2}{w_y^2(z)}\right],
\end{equation}
with
\begin{equation}\label{Elaser2}
    w_y(z)=w_0\sqrt{1+\left(\frac{z}{z_R}\right)^2}, \;|E(z)|=|E_0|\frac{w_0}{w_y(z)}, 
\end{equation}
where $z_R=\pi w_0^2/\lambda$ is the Rayleigh range, $\lambda$ being the field wavelength inside the medium \cite{mandel}. {We consider the finite longitudinal extent of the laser pulse in the $z$ direction, $L_p=cT/n$, where $n$ is the refractive index of the crystal. Thus, only the portion of the crystal instantaneously illuminated by the pulse contributes to the induced polarization.}

It is important to note that even though the beam is focused between the two holes, a non-zero polarization is present in the hole walls which, in turn, creates a non-zero average field inside the holes. That field is, however, small compared to the field between the holes, since around the circular slits (when $z\approx 0$), the rectified field intensity falls with a factor of $\exp\left[-{2y^2}/{w_0^2}\right]$.

If we consider that the laser beam is sufficiently collimated, so that the induced surface charges in the holes are sufficiently small and do not influence the polarization inside the crystal,  according to Eqs. (\ref{ORfield}) and (\ref{Elaser1}) the average polarization in the region between the holes  is given by
\begin{equation}
\mathbf{P}_{OR} = \varepsilon_0\chi^{(2)}|E(z)|^2 \exp\left[-\dfrac{2y^2}{w_y^2(z)}\right] \hat{\mathbf{y}}. \label{pol}
\end{equation}

\begin{figure}
  \centering
  \includegraphics[width=0.4\textwidth]{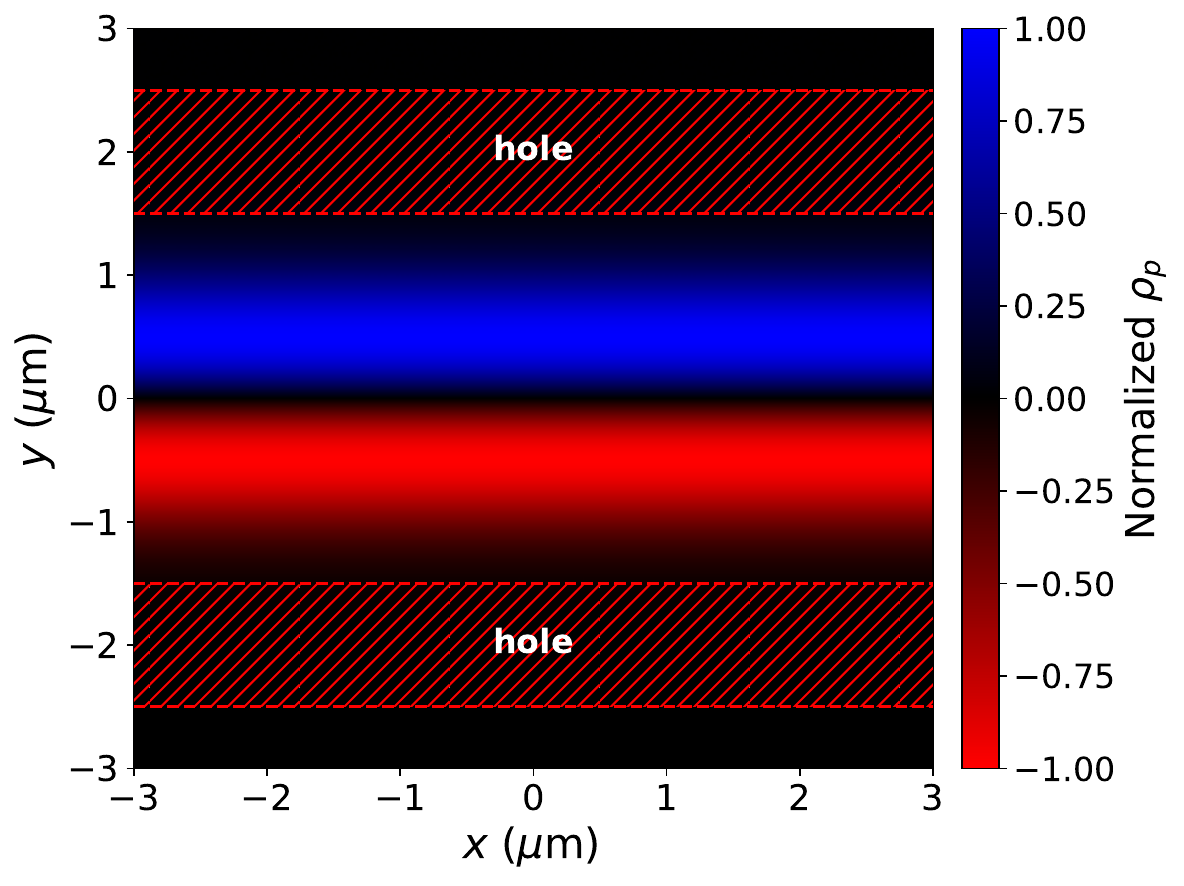}  
  \caption{Bulk  polarization charge density $\rho_p$ from Eq. (\ref{rhop}) induced in the crystal between the holes, for $w_0$ = 1 $\mu$m and a separation $d=3\mu$m between the holes.}
  \label{charges}
\end{figure}

The volumetric $\rho_p = -\nabla\cdot\mathbf{P}$, and surface $\sigma_p = \mathbf{P}\cdot\hat{\mathbf{n}}$ polarization charge densities induced by the laser are then given by
\begin{eqnarray}
\rho_p &=& \varepsilon_0\chi^{(2)}|E(z)|^2\dfrac{4y}{w_y^2(z)}\exp\left[-\dfrac{2y^2}{w_y^2(z)}\right], \label{rhop}\\
\sigma_p &=& \pm\varepsilon_0\chi^{(2)}|E_0|^2\exp\left[-2\dfrac{(y_c-R\cos\theta)^2}{w_0^2}\right]\cos\theta, \nonumber \\\label{sigmap}
\end{eqnarray}
where the + (-) sign of Eq. (\ref{sigmap}) is taken for the hole centered at $y=y_c$ ($y=-y_c$), $R$ is the radius of both holes, and $\theta$ is the polar angle. The angle $\theta = 0$ is taken so that these are the points closer to the center of the beam. This means that, for the hole centered in $+y_c$ ($-y_c$), $\theta =0$ corresponds to the lower (upper) edge of the hole. 
In Fig. \ref{charges} we see the bulk charge distribution $\rho_p$ in between the holes. 

The rectified electric field $\mathbf{E}_{OR}$ can be decomposed into $\mathbf{E}_{OR} = \mathbf{E}_{\rho}+\mathbf{E}_{\sigma}$, where $\mathbf{E}_{\rho}$ ($\mathbf{E}_{\sigma}$) is the electric field created by the bulk (surface) polarization charges. These fields can be computed using Coulomb's law:
\begin{equation}\label{erho}
\mathbf{E}_{\rho}(\mathbf{r}) = \dfrac{1}{4\pi\varepsilon_0}\int \rho_p(\mathbf{r}^\prime)\dfrac{\mathbf{r}-\mathbf{r}^\prime}{|\mathbf{r}-\mathbf{r}^\prime|^3}dV^\prime,
\end{equation}
\begin{equation}\label{esigma}
\mathbf{E}_{\sigma}(\mathbf{r}) = \dfrac{1}{4\pi\varepsilon_0}\int \sigma_p(\mathbf{r}^\prime)\dfrac{\mathbf{r}-\mathbf{r}^\prime}{|\mathbf{r}-\mathbf{r}^\prime|^3}da^\prime,
\end{equation}
with $\rho_p$ and $\sigma_p$ given by Eqs. (\ref{rhop}) and (\ref{sigmap}).  { For a finite laser pulse, the polarization charges contributing to
Eqs.~\eqref{erho} and \eqref{esigma} vary as the pulse propagates
through the crystal. We therefore calculate the electric fields as
their temporal averages during the interval in which the pulse crosses
the plane $z=0$, from the arrival of its leading edge until the
departure of its trailing edge.}

\begin{figure}
    \centering
  \begin{minipage}[b]{0.4\textwidth}
    \centering
    \includegraphics[width=\textwidth]{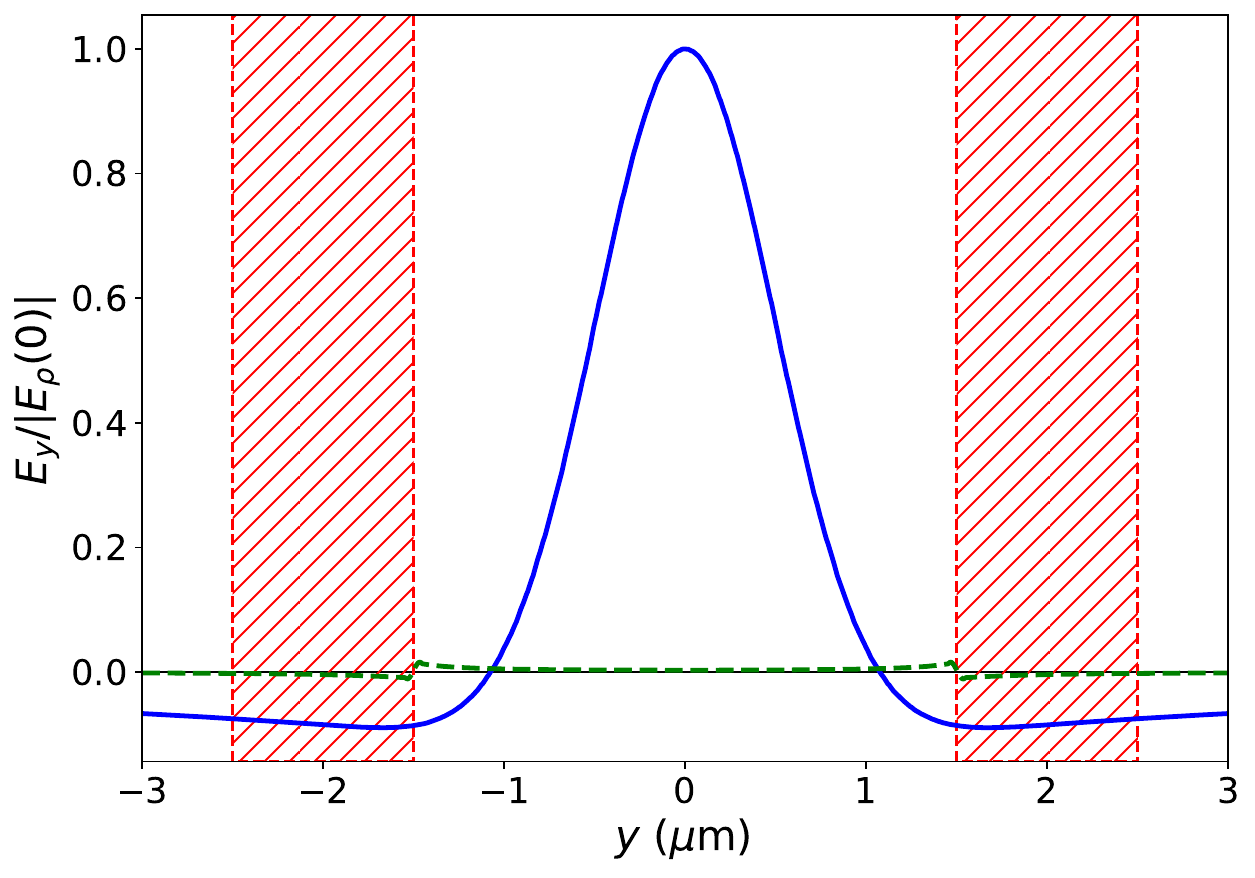}
    \par\centering (a)
  \end{minipage}
  \begin{minipage}[b]{0.4\textwidth}
    \centering
    \includegraphics[width=\textwidth]{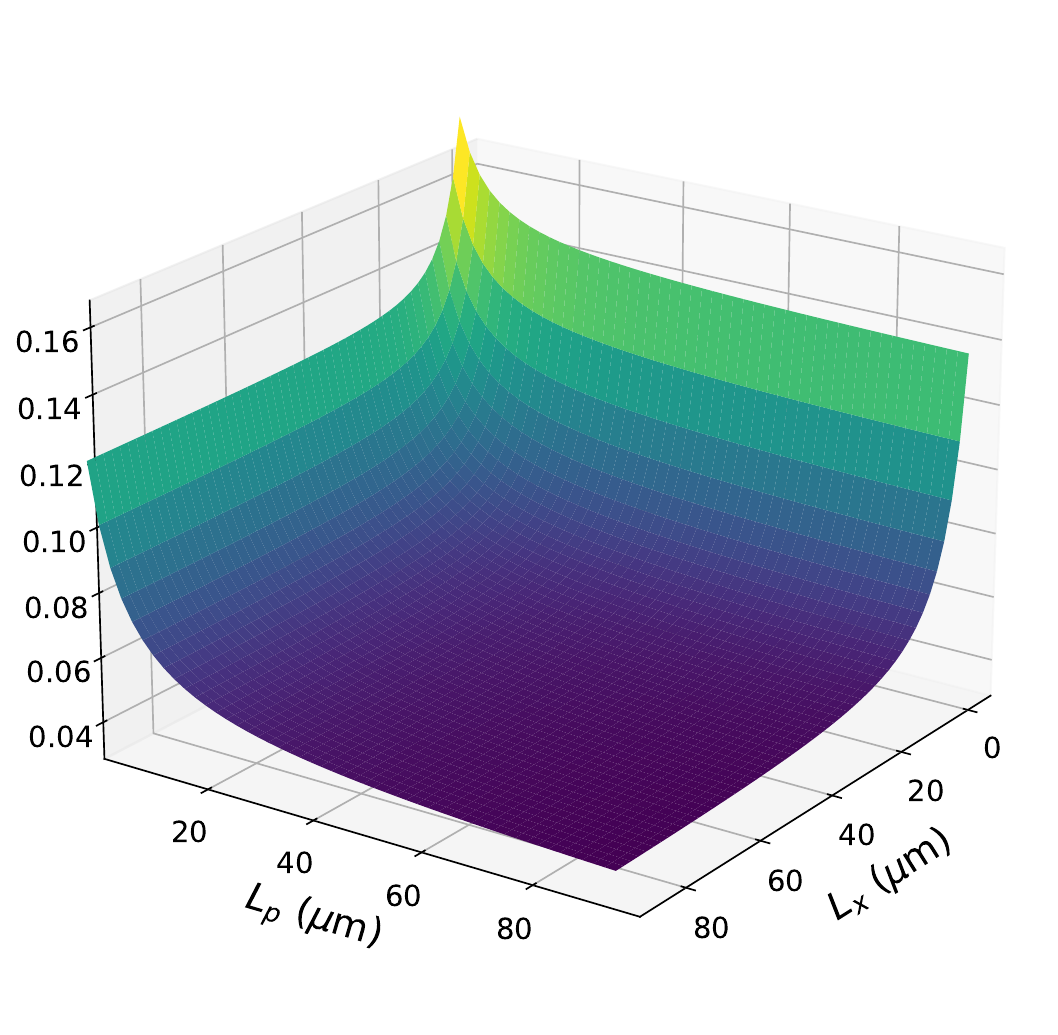}
    \par\centering (b)
  \end{minipage}
    \caption{(a) $y$-component of the time-averaged electric fields generated the bulk polarization charges $\mathbf{E}_\rho$ (blue continuous line) and surface polarization charges $\mathbf{E}_\sigma$ (green traced line), as a function of $y$ near the center of the laser beam, at $x=0$ and $z=0$.  The fields were computed from  Eqs. (\ref{erho}) and (\ref{esigma}) with the parameters $d=3 \mu$m, $w_0 = 1$ $\mu$m, $L_x = 20$ $\mu$m, and $T=100$ fs ($L_p=17\mu$m ). (b) Relative intensity of residual fields inside the holes as a function of the slab width $L_x$ {and of the pulse length $L_p=cT/n$,} keeping the other parameters fixed.}
    \label{EFields}
\end{figure}

Since $\mathbf{E}_{\sigma}$ depends on the surface charges $\sigma_p$ induced at the holes edges by the laser beam, it can be very small for focused beams, with a small amplitude at the holes edges. Considering $\mathbf{E}_{\rho}$, if the beam width in the $y$ direction is much smaller than its width in the $x$ direction and also much smaller than its width in the $z$ direction, as we assume here, the charge distribution $\rho_p$ is like the one depicted in Fig. \ref{charges}. That charge distribution resembles the one of a parallel plate capacitor with the separation between its plates being much smaller than the dimensions of the plates. In this way, we expect that the electric field between the holes, which will be responsible for an AB phase difference, be much larger than the fields at the holes, which interact locally with the electrons in an AB scheme.  Fig. \ref{EFields}, constructed by numerical calculations, displays the $y$-component of the electric field obtained from Eqs. (\ref{erho}) and (\ref{esigma}) in function of $y$  at $x=0$ and $z=0$, for holes of diameter $2R=1\mu$m, holes separation {$d=3 \mu$m}, beam waist $w_0 = 1$ $\mu$m,  crystal width $L_x = 20\mu$m and pulse length $L_p=17\mu$m (pulse duration $T=100$ fs). As it can be seen in Fig. \ref{EFields}, the field generated by the surface charges $\sigma_p$ is negligible compared to that created by the bulk charges $\rho_p$ and, therefore, it will be neglected in future calculations.

In the proposed experiment, electrons propagating in the $x$-direction are diffracted by the crystal slab with two holes carved on it, as depicted in Fig. \ref{setup}. The electrons are sent to the interferometer one by one and form an interference pattern in a screen positioned in a distance $D$ from the crystal, far away. When each electron is in a superposition state passing through the two holes, a pulsed laser beam induces the bulk charge density from Eq. (\ref{rhop}), depicted in Fig. \ref{charges}, which generates the electric field from Eq. (\ref{erho}) in the system. We assume that the electron moves a negligible amount while the laser pulse interacts with the slab generating the induced charge. The electric field between the two holes induces a potential difference between the paths, changing the interference pattern in an electric AB effect. 

The AB phase difference for an electron in a superposition of two paths $a$ and $b$ is given by \cite{aharonov59}
\begin{eqnarray}
    \phi_{AB} &=& -\dfrac{e}{\hbar}\int\left(V(\mathbf{r}_a)-V(\mathbf{r}_b)\right)dt \label{phiAB1},
\end{eqnarray}
where $V(\mathbf{r}_a)$ ($V(\mathbf{r}_b)$) is the electric potential in path $a$ ($b$). We can rewrite Eq. (\ref{phiAB1}) as
\begin{eqnarray}
    \phi_{AB} &=& \dfrac{e}{\hbar}\int_0^T \int_{\mathbf{r}_a}^{\mathbf{r}_b}\mathbf{E}\cdot d\mathbf{r} \;dt \label{phiAB2},
\end{eqnarray}
$\mathbf{E}$ being the system electric field. We can use Eq. (\ref{phiAB2}) to numerically calculate the AB phase created as the electron traverses the holes due to the field from Eq. (\ref{erho}). During the passage of the laser pulse through the crystal, the electron which is in a superposition between the center of both holes experience the {time-averaged } fields shown in Fig. \ref{EFields} (a). { If we consider a KTP slab ($\chi^{(2)} \approx 33 \times 10^{-12}$ m/V), with the geometric parameters mentioned above and a 100 fs laser pulse, we find that we need a laser intensity $I_0 = n\varepsilon_0c|E_0|^2/2 \approx 4.3 \times10^8$ W/cm$^2$  in order for the AB phase to be $\phi_{AB} \approx \pi$. In this case, the maximum value of the induced electric field, at the origin of the system of coordinates, is around  $2.2\times10^4$ V/m,  and the field inside the holes is 10 times smaller.} By changing the laser intensity from 0 to $I_0$ in various steps, the positions of the maxima and minima of the electrons interference pattern would change, demonstrating the electric AB effect.

\begin{figure}
  \centering
  \begin{minipage}[b]{0.4\textwidth}
    \centering
    \includegraphics[width=\textwidth]{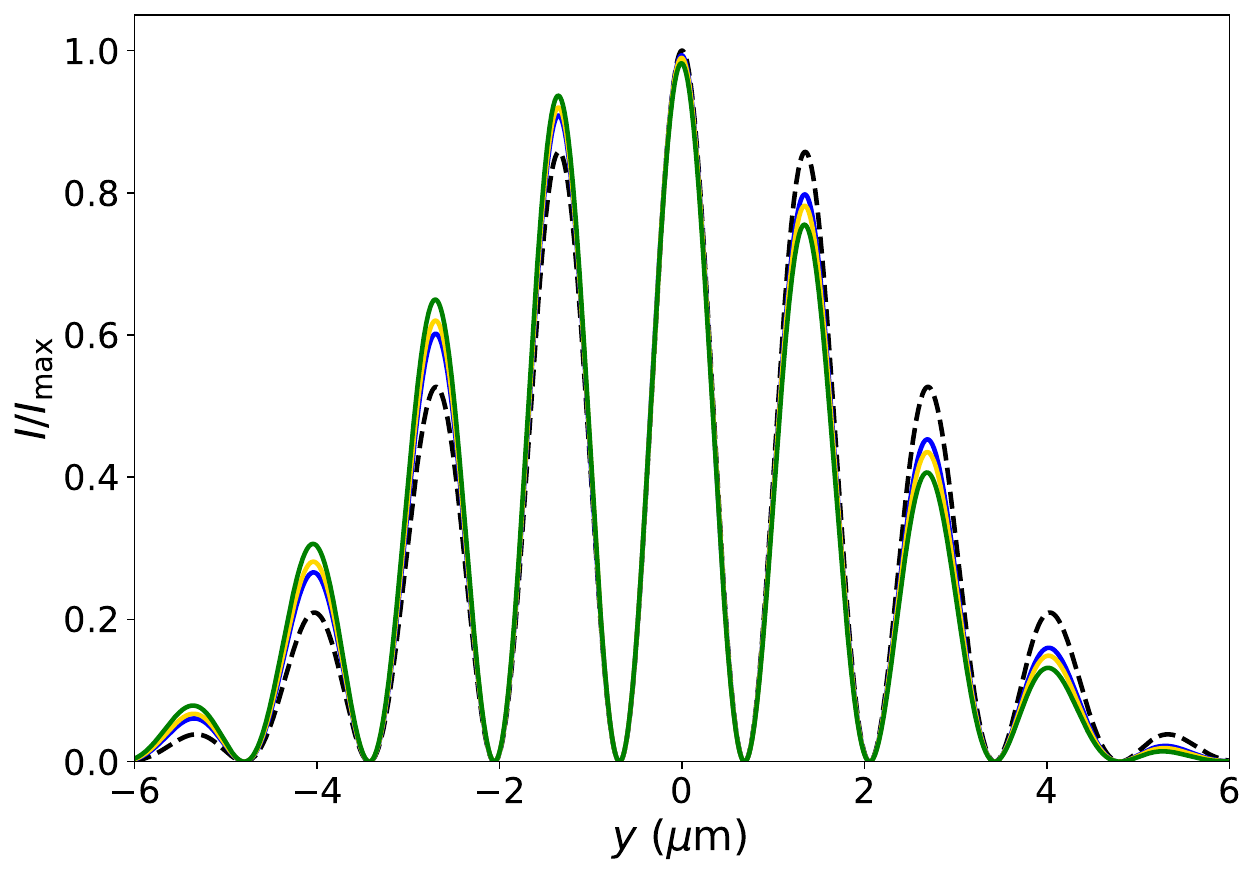}
    \par\centering (a)
  \end{minipage}
  \begin{minipage}[b]{0.4\textwidth}
    \centering
    \includegraphics[width=\textwidth]{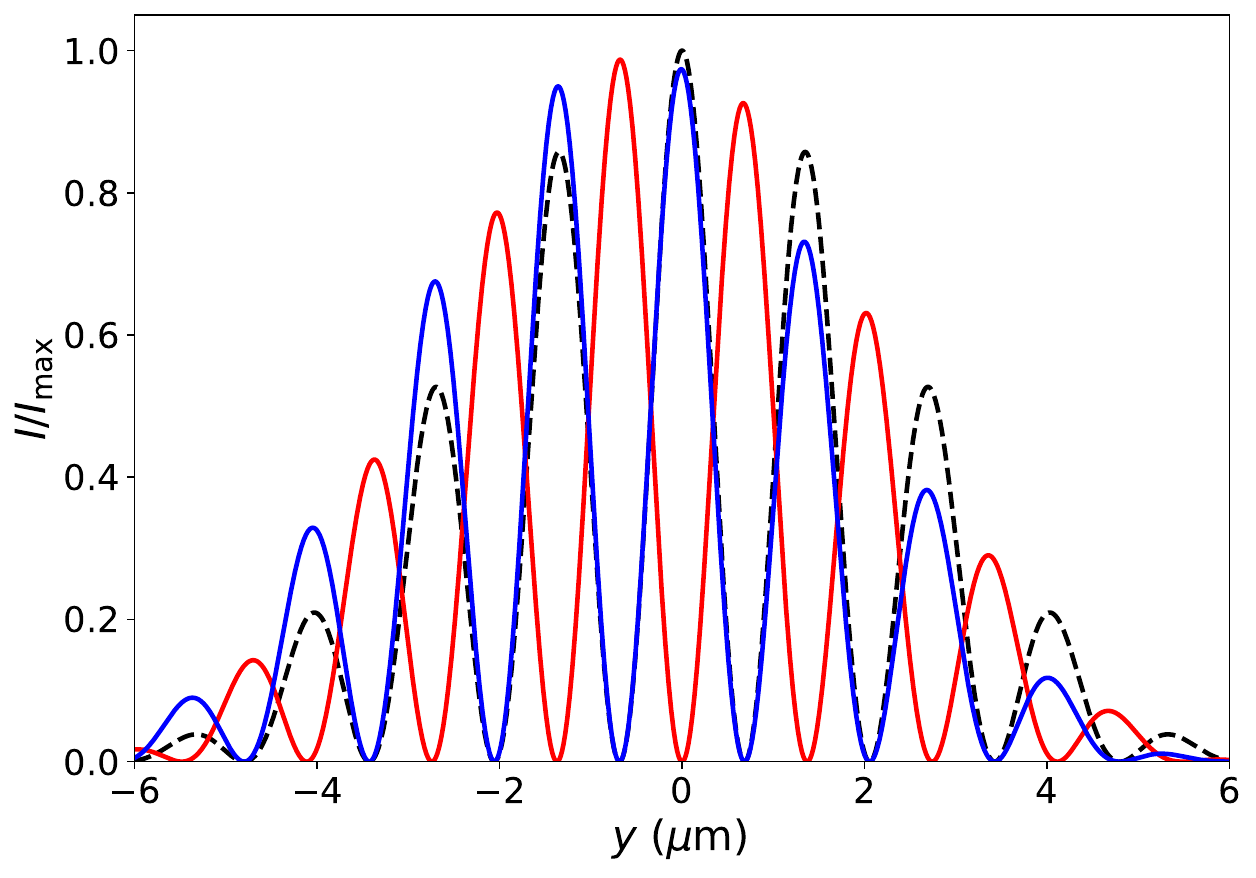}
    \par\centering (b)
  \end{minipage}
    \caption{ Interference patterns calculated within the paraxial approximation, $k_y=k_e y/D$, with $\tilde{\psi}_0(k_y)\propto 2J_1(k_yR)/(k_yR)$ for a circular aperture. In both panels, $D=10$ cm, $T=100$ fs, $d=3~\mu$m  $w_0=1~\mu$m $k_e=1.15\times10^{11}$ m$^{-1}$, corresponding to 500 eV electrons. (a) Effect of the residual field for different slab widths $L_x$, excluding the AB phase. The curves, obtained from Eq.~(\ref{Intensity_kick}), correspond to $\delta k=0$ (black), $L_x=20~\mu$m (blue), $10~\mu$m (yellow), and $5~\mu$m (green). The laser intensity is $I_0=3.5\times10^8$ W/cm$^{2}$, corresponding to $\phi_{\rm AB}=\pi$. (b) Full interference pattern including both the dynamical momentum shift due to the residual field and the AB phase,   with $L_x=20\mu$m. The dashed black curve corresponds to $I_0=0$, while the solid red, and blue curves correspond, respectively, to $I_0=4.3\times10^8$, and $8.6\times10^8$ W/cm$^{2}$, yielding $\phi_{\rm AB}=\pi$, and $2\pi$.}
    \label{fields}
\end{figure}

At the same time, however, the electron will be subject to an effectively uniform field within the holes, as can be seen in Fig. \ref{EFields} (a).  
 The intensity of the residual field inside the holes decreases with the increase of the width of the crystal slab in the $x$-direction $L_x$ and with the increase of the pulse length in the $z$ direction $L_p$, as can be seen in Fig. \ref{EFields}(b). As the electrons traverse through the holes, the local interaction with the residual field results in a transverse momentum transfer $\Delta p_y=\hbar\delta k$, producing a small deflection angle $\theta$. On Fig. \ref{fields}, we show the effects of this local interaction with the non-zero fields on the interference pattern, on panel (a), as well as the full interaction picture, including the effects of both the local interaction and the AB phase, on panel (b). On these plots, the calculated values for the deflection angles were of the order $\theta\approx 10^{-6}$ rad. In the Appendix the construction of the curves from Fig. \ref{fields} are explained. As it can be seen from Fig. \ref{fields} (a), the effect of the residual field on the interference pattern is to shift the envelope function of the pattern, while keeping the position of the fringes intact.  This can be understood as follows. Since the residual electric field inside both holes is essentially identical, the electron wavefunctions traversing the two paths acquire the same momentum kick while crossing the slab. Moreover, this momentum transfer is several orders of magnitude smaller than the initial electron momentum and is predominantly transverse to its propagation direction. As a result, the electron wavelength remains practically unchanged. Since the momentum kick includes no relative phase between the electron wavefunctions at each hole, the relative phase responsible for the interference fringes is preserved. The only effect of the residual field is a slight deflection of the electron trajectory, which modifies the overall intensity envelope but does not change the locations of the interference maxima and minima. {The full interference pattern for various laser intensities $I_0$ and, therefore, various values for the AB phase, can be seen in Fig. \ref{fields} (b).} Since the electric AB effect shifts the fringe positions, whereas the residual field leaves them unchanged, the fringe displacement observed as the laser intensity is varied can be unambiguously attributed to the electric AB phase. 

{

}

\section{Conclusion}\label{conc}

In this work, we have proposed an experimentally accessible scheme to observe an optically induced electric Aharonov-Bohm effect. In our proposal, a short laser pulse propagating through a nonlinear crystal generates, through optical rectification, a transient quasi-DC polarization field between the two holes of an electron double-slit interferometer. An electron coherently passing through the two holes can therefore acquire a controllable relative phase even though the electric field at the electron paths is strongly suppressed. By varying the laser intensity, the resulting Aharonov-Bohm phase can be continuously tuned, producing a corresponding displacement of the electron interference fringes.

For realistic parameters, considering a $100\,\mathrm{fs}$ laser pulse, we find that an Aharonov-Bohm phase of order $\pi$ can be obtained with laser intensities of order $10^{8}\,\mathrm{W/cm^2}$. Although the finite size of the crystal and the polarization charges at the apertures generate residual electric fields inside the holes, these fields are substantially weaker than the field in the region separating the two paths. More importantly, their effect on the interference pattern is qualitatively distinct from that of the Aharonov-Bohm phase: the residual field produces a common transverse momentum kick to the two wave packets and therefore shifts the diffraction envelope without changing the positions of the interference fringes, whereas the Aharonov-Bohm phase directly shifts the fringes themselves. This difference provides a clear experimental signature by which the phase effect can be distinguished from local forces acting on the electrons.

The proposed scheme therefore provides a route toward observing the electric Aharonov-Bohm effect in a regime where local electric-field effects on the interfering electrons are strongly suppressed and can be independently identified. The use of optical rectification also makes the induced field intrinsically transient and externally controllable, allowing the electric field to be switched on only while the electron is coherently distributed between the two paths. Beyond providing a possible experimental test of the electric Aharonov-Bohm effect, this approach illustrates how nonlinear optical control may be used to engineer electromagnetic phases in free-electron interferometry.

\section*{Acknowledgements}
GPM and PLS are grateful to the funding of the Conselho Nacional de Desenvolvimento Científico e Tecnológico (CNPq), specially through Grants No. 446612/2024-3 and 302881/2025-5. HB was supported by the National Science Foundation under Grant No. PHY2513170.

\appendix

\section{Interference pattern in the presence of the residual electric field}
\label{app:interference}

In this Appendix, we derive the interference patterns shown in Fig. \ref{fields}. While the electron is inside the crystal, its wavefunction is given by the coherent superposition of the wavefunctions on the upper and lower holes. Since the holes are identical, the two wavefunctions differ only by a translation along the $y$ direction, and we have
\begin{equation}
    \psi(y,z)= \psi_0(y-y_c,z)+\psi_0(y+y_c,z),
\end{equation}
where $\pm y_c$ are the coordinates of the centers of the two holes.

As discussed in the main text, the residual electric field inside the holes is approximately uniform, as seen in Fig. \ref{EFields} (a). During a laser pulse of duration $T$, the electron receives the transverse momentum $\Delta p_y=
eE_{\rm hole}T$, corresponding to the wavevector kick $\delta k=\Delta p_y/\hbar=eE_{\rm hole}T/\hbar.$ Since both electron wave packets experience essentially the same electric field, they receive the same transverse momentum kick. In this way, the electron wavefunction becomes
\begin{equation}\label{psi_kick}
    \psi(y,z)= \psi_0(y-y_c,z)e^{i\delta k(y-y_c)}+\psi_0(y+y_c,z)e^{i\delta k(y+y_c)}.
\end{equation}
The momentum kick generates no relative phase between the holes, such that the phases at the holes are equal.

In the Fraunhofer limit, the electron wavefunction at the detection screen is proportional to the Fourier transform of the wavefunction immediately after the crystal \cite{Goodman2017}: 
\begin{equation}
\tilde\psi(k_y,k_z)
=
\int
dy\,dz\,
\psi(y,z)
e^{-i(k_yy+k_zz)}.
\label{fourier_definition}
\end{equation}
Using Eq.~(\ref{psi_kick}), we obtain
\begin{equation}
\tilde\psi(k_y,k_z)
=
\tilde\psi_0(k_y-\delta k,k_z)
\left(
e^{-ik_yy_c}
+
e^{ik_yy_c}
\right).
\end{equation}
The corresponding intensity distribution is
\begin{equation}
I(k_y,k_z)
=
4
\left|
\tilde\psi_0(k_y-\delta k,k_z)
\right|^2
\cos^2(k_yy_c).
\label{Intensity_kick}
\end{equation}

To express the interference pattern in terms of the spatial coordinates on the detection screen, we consider a screen at a distance $D$ from the holes. In the paraxial approximation, the transverse components of the electron wavevector are related to the coordinates $(y_s,z_s)$ on the screen by
$k_y \simeq k{y_s}/{D}$, and 
$k_z \simeq k{z_s}/{D}$,
where $k=2\pi/\lambda_e$ is the magnitude of the electron wavevector. Therefore, the intensity distribution in real space is obtained from Eq.~(\ref{Intensity_kick}) by the substitutions $k_y\rightarrow ky_s/D$ and $k_z\rightarrow kz_s/D$. The spatial scale of the interference pattern consequently increases linearly with the distance $D$ between the holes and the detection screen.

Equation~(\ref{Intensity_kick}) shows that the residual electric field affects only the diffraction envelope through the factor $|\tilde\psi_0(k_y-\delta k,k_z)|^2$, which is translated in momentum space by the amount $\delta k$. In contrast, the interference term $\cos^2(k_yy_c)$ is completely independent of $\delta k$. Consequently, the positions of the interference maxima and minima remain unchanged, while the diffraction envelope is displaced. The interference patterns shown in Fig.~\ref{fields}(a) were obtained from Eq.~(\ref{Intensity_kick}) using the values of $\delta k$ calculated for each slab width.

For comparison, if the electrons acquire an electric AB phase difference $\phi_{AB}$ while traversing the interferometer, the wavefunction in momentum space becomes
\begin{eqnarray}
\tilde\psi(k_y,k_z)
&=&
\tilde\psi_0(k_y-\delta k,k_z)
\left[
e^{-ik_yy_c}
+
e^{i\phi_{AB}}
e^{ik_yy_c}
\right],\nonumber \\
\label{FT_AB}
\end{eqnarray}
which gives
\begin{equation}
I(k_y,k_z)
=
4
\left|
\tilde\psi_0(k_y-\delta k,k_z)
\right|^2
\cos^2
\left(
k_yy_c+\frac{\phi_{AB}}{2}
\right).
\label{Intensity_AB}
\end{equation}
Unlike the momentum kick, the AB phase changes the relative phase between the two electron wave packets. Consequently, it shifts the interference fringes themselves rather than the diffraction envelope.  The interference patterns shown in Fig.~\ref{fields}(b) were obtained from Eq.~(\ref{Intensity_AB}).

\bibliography{bibliography}

\end{document}